\documentclass[pre,aps,reprint,amsmath,amssymb,superscriptaddress]{revtex4-1}

\usepackage{color}
\usepackage{amssymb}
\usepackage{esint}

\usepackage{amstext}
\usepackage{graphicx,epstopdf}
\usepackage{color}
\usepackage{amsmath}
\usepackage{color}
\usepackage{enumitem}

\makeatletter

\begin{document}

\title{Nonergodic subdiffusion from transient interactions with heterogeneous partners}

\author{C. Charalambous}
\author{G. Mu\~noz-Gil}
\author{A. Celi}
\affiliation{ICFO-Institut de Ci\`encies Fot\`oniques, The Barcelona Institute of Science and Technology, 08860 Castelldefels (Barcelona), Spain}
\author{M.F. Garcia-Parajo}
\author{M. Lewenstein}
\affiliation{ICFO-Institut de Ci\`encies Fot\`oniques, The Barcelona Institute of Science and Technology, 08860 Castelldefels (Barcelona), Spain}
\affiliation{ICREA - Pg. Llu\'{\i}s Companys 23, 08010 Barcelona, Spain}
\author{C. Manzo}
\email{carlo.manzo@uvic.cat}
\affiliation{ICFO-Institut de Ci\`encies Fot\`oniques, The Barcelona Institute of Science and Technology, 08860 Castelldefels (Barcelona), Spain}
\affiliation{Universitat de Vic - Universitat Central de Catalunya (UVic-UCC), C. de la Laura,13, 08500 Vic, Spain }
\author{M.A. Garc\'{\i}a-March}
\email{miguel.garcia-march@icfo.es}
\affiliation{ICFO-Institut de Ci\`encies Fot\`oniques, The Barcelona Institute of Science and Technology, 08860 Castelldefels (Barcelona), Spain}



\begin{abstract}
Spatiotemporal disorder has been recently associated to the occurrence of anomalous nonergodic diffusion of molecular components in biological systems, 
but the underlying microscopic mechanism is still unclear. We introduce a model in which a particle performs continuous 
Brownian motion with changes of diffusion coefficients induced by transient molecular interactions with diffusive binding partners. 
In spite of the exponential distribution of waiting times, the model shows subdiffusion and nonergodicity similar to the heavy-tailed continuous time random walk. 
The dependence of these properties on the density of binding partners is analyzed and discussed. 
Our work provides an experimentally-testable microscopic model to investigate the nature of nonergodicity in disordered media.
\end{abstract}

\maketitle
\section{Introduction}

Advances in fluorescence-based videomicroscopy have recently enabled tracking the movement of small particles and individual molecules with nanometer precision in living cells~\cite{2015ManzoROPP}. 
These experiments have allowed the evaluation of several observables which characterize the dynamics of a variety of biomolecules. 
Remarkably, in many biological systems, they have revealed that the mean squared displacement (MSD) shows a nonlinear growth in time
\begin{equation}
\label{eq:anomalous}
\mbox{MSD}(t) \sim t^\alpha \,\,\,\, \mbox{with}  \, 0<\alpha<1,
\end{equation}
and thus deviates from Brownian motion ($\alpha=1$). 

Such anomalous diffusion can have different physical origins and a large amount of models have been proposed for its interpretation, some of the most relevant ones being recently discussed by Metzler and co-workers~\cite{2014MetzlerPCCP}.
The detailed analysis of the particle trajectories has shown that some processes characterized by {\it anomalous} diffusion also exhibit differences between ensemble and time averaged observables, 
such as the mean squared displacement itself~\cite{2006GoldingPRL, 2011JeonPRL, 2011WeigelPNAS, 2013TabeiPNAS, 2015ManzoPRX}. 
This feature, known as weak-ergodicity breaking~\cite{1992Bouchaud}, reflects the physical nature of specific stochastic mechanisms, 
for which time averages are random and thus irreproducible in spite of the large statistics. 
These properties are captured by a number of nonstationary stochastic processes~\cite{2014MetzlerPCCP}, 
such as the heavy-tailed continuous time random walk (CTRW)~\cite{1975ScherPRB}, or heterogeneous diffusion processes~\cite{2014MetzlerPCCP}.

All these models postulate the presence of some form of kinetic heterogeneity in the system,  e.g. a nonconstant diffusion coefficient, 
which causes the nonergodic behavior. Recent measurements reporting on the variation of diffusion coefficient in space or time~\cite{2008ManleyNatMet, 2013CutlerPlosONE, 2014MassonBJ}, 
sometimes associated with anomalous and nonergodic diffusion~\cite{2015ManzoPRX}, justify these assumptions phenomenologically.  
However, some questions remain unanswered: What causes such kinetic heterogeneity at the microscopic level? 
Which is the associated mechanism leading to nonergodic anomalous diffusion? 
Experimental evidences have recently shown that specific interactions between molecules diffusing on the cell membrane can temporarily modify their diffusive behavior~\cite{2015TorrenoJPD} 
but there is lack of a microscopic model encompassing a mechanistic link between interactions and nonergodic subdiffusion.

In this paper, we introduce a model of ordinary CTRW in which anomalous diffusion and nonergodicity arise as a consequence of transient molecular interactions with a population of diffusing binding partners. 
In our simple physical picture, the Brownian motion of a random walker called {\it prey}, 
is temporarily modified upon interactions with a set of random walkers called {\it hunters}: as a consequence of the interaction, 
the motion of the prey is altered by a factor randomly drawn from a distribution with a power law tail. 
A particular case of this general scenario is provided by a distribution obtained considering that the diffusivity is the sum of several squared Gaussian random variables arising out of many microscopic degrees of freedom~\cite{2001BeckPRE}. We also consider a case in which interactions slow down the diffusion of the prey. From the biophysical point of view, the model reflects the scenario of a diffusing molecule which, upon interaction with another molecule, undergoes a transient change of diffusivity for the time the interaction takes place. An example of this effect is provided by the formation of transient dimers and oligomers among chemically identical molecules~\cite{2011DLidke, 2012Kusumi}, as well as by interactions occurring among different molecular species. Since these scenarios might include a variety of interactions between a multiplicity of cellular components (such as proteins, lipids, vesicles, or organelles), for the sake of generality we will refer to the random walkers as {\it prey} and {\it hunters}.  

The broad diffusivity distribution of our model is in agreement with several experimental reports~\cite{1997SaxtonBJ}. In addition, it accounts for the presence of heterogeneity in interaction products (i.e., formation of clusters with a varying number of molecules and thus different diffusivities) as well as in interacting partners (i.e., interactions with a pool of chemically different species with different diffusivity) that can be found in the cellular environment. 

As for the heavy-tail CTRW, also in this case the ergodicity breaking is generated by energetic disorder due to transient chemical binding~\cite{1975ScherPRB}. 
However, in the model discussed here, the waiting times are always drawn from an exponential distribution; therefore, even while interacting, the prey always perform Brownian motion although with different diffusivities.  Following the classification given in Ref.~\cite{2014MetzlerPCCP}, our model thus describe a heterogeneous diffusion process. 
A crucial aspect of this model is that we explicitly introduce a physical mechanism (i.e., transient interactions) causing the change of diffusivity. Moreover, even though the prey experiences the disorder only for a limited amount of time (i.e., upon interaction), this is sufficient to generate anomalous diffusion and nonergodicity even in the dilute limit.  
The occurrence of such interactions can be experimentally verified, e.g. using multicolor single particle tracking experiments~\cite{2015TorrenoJPD}. In addition, the model allows one to estimate microscopic parameters of the system under investigation. This sets a great advantage over models with diffusion coefficient varying in space~\cite{2013CherstvyPCCP} or time~\cite{2014MassignanPRL}. After describing the model, we calculate the effective waiting time distribution and the ensemble-averaged MSD (eMSD), discussing the origin of subdiffusivity. 
Next, we present the results of numerical simulations, for which we also calculate the time-averaged MSD (tMSD) and the ergodicity breaking (EB) parameter. 
Last, we discuss potential applications and address future developments of the model.


\begin{figure}
\includegraphics[width=0.95\columnwidth]{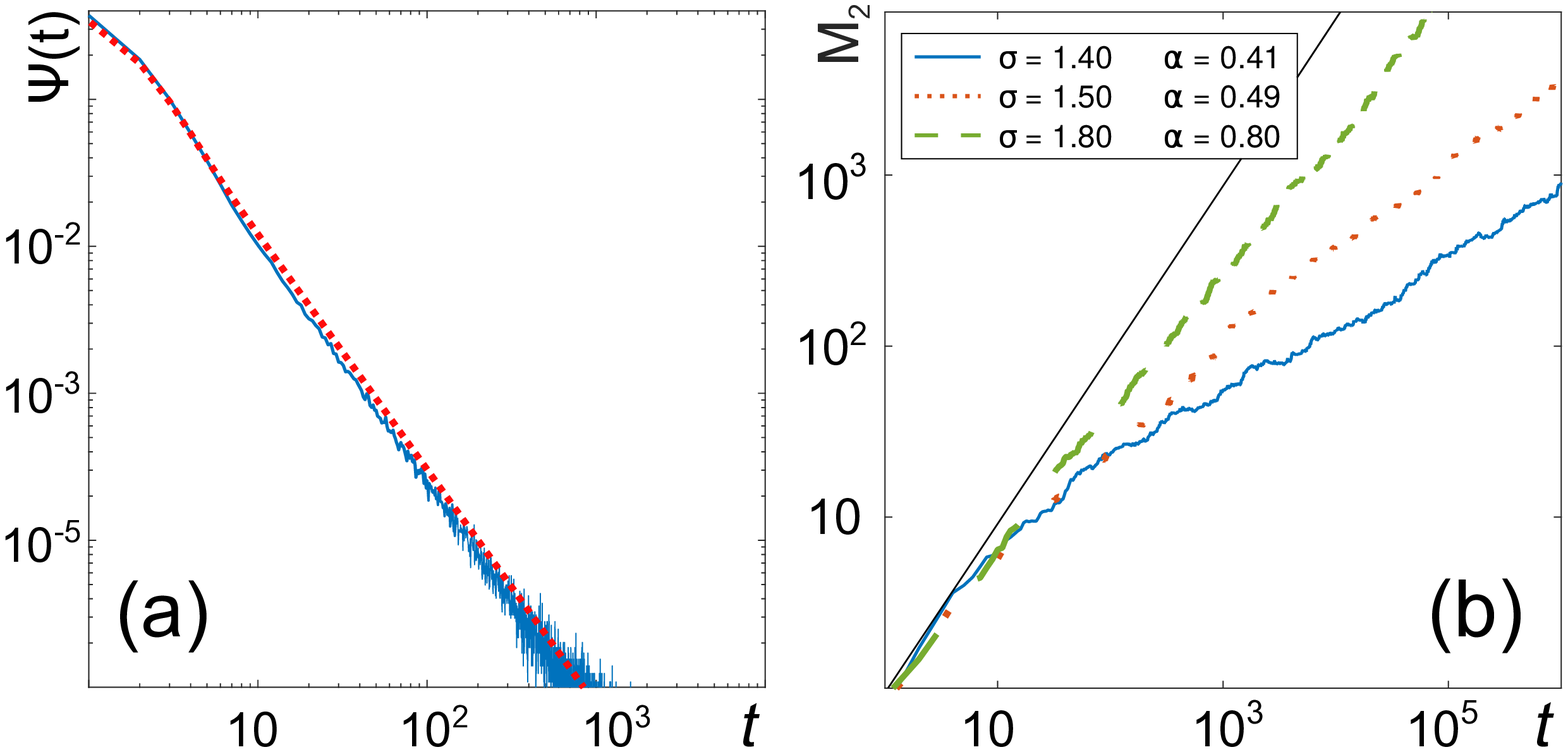}
\caption{(a) Analytical (dashed line) vs numerically-calculated (solid line) waiting time distribution for $\sigma=1.6$. 
(b) Ensemble mean squared displacement calculated for different values of $\sigma$, showing the subdiffusive behavior in the asymptotic regime and the agreement of the exponent  $\alpha$  with the theoretical predicted values. 
The black solid thin line indicates the exponent $\alpha=1$. These results correspond to $N=8$ hunters and $L=20$ sites.   \label{fig1} }
\end{figure}

\section{The model}

We consider a random walker ({\it prey}) interacting with a system of $N$ independent random walkers ({\it hunters}) moving in $d$-dimensional regular lattices with $m=L^d$ sites and $d=1,2$. Periodic boundary conditions ensure constant hunters' density in time.
Both types of walker perform a CTRW with waiting times drawn from an exponential distribution $\psi_0(t)=(1/\tau)\exp[-t/\tau]$. 
The interaction between the prey and the hunters is devised as follows: when the prey coincides at a lattice site with at least one of the hunters, 
the waiting time for the next step is drawn from a distribution  
 $\psi_\kappa(t)=(1/\kappa\tau)\exp[-t/\kappa\tau]$, with $\kappa$ stochastically chosen from any probability distribution $P_{\kappa}\left(\kappa\right)$, such that as $k\rightarrow \infty$
\begin{equation}
 \begin{array}{rl}
 \label{eq:Pkappageneral}
P_{\kappa}\left(\kappa\right) \approx \kappa^{-\sigma}  & \text{with } \sigma>1,
\end{array}
\end{equation}
and with $P_{\kappa}\left(\kappa\right)$ decaying rapidly to $0$ for small $\kappa$. Every time a hunter and the prey meet, a new $\kappa$ is randomly drawn; therefore, we will refer to this process as  {\it annealed} disorder. Alternatively, we can formulate a similar model by assigning a different value of $\kappa$ to every hunter with the same probability as in Eq.~(\ref{eq:Pkappageneral}). 
In this case, every hunter maintains the same $\kappa$ until it leaves the lattice, and assumes a new value at reentering.  The latter case represents a {\it quenched} disorder of the binding rates $\kappa$.  
We will show that for large enough densities,  the probability that the prey reencounters the same hunter before hitting a large number of other hunters is negligibly small; therefore, a different value of $\kappa$ is sampled at every interaction, thus establishing the equivalence between both the annealed and quenched realizations of the model. 

The probability for the prey to perform $n$ steps with waiting times $t_i$ in a time $t=\sum_{i=1}^n t_i$ is
\begin{equation*}
 \psi_n(t)\!=\!\int_0^t\!\!\!\dots\!\int_0^t\!\! \psi(t_1)\dots  \psi(t_n)\delta\!\left(t-\sum_{i=1}^n t_i\right)dt_1\dots dt_n,
\end{equation*}
where the waiting time distribution $  \psi (t)$ must account for the cases of encountering [$  \psi (t) = \psi_\kappa$] or not a hunter [$  \psi (t) =\psi_0$]. 
If $p_{\rm{NH}}$ is the probability of not hitting a hunter, and $p_{\rm{H}}=1-p_{\rm{NH}}$ its complement, we can then write
\begin{eqnarray}
\label{Eq:Waitingtime}
&  \psi (t)= p_{\rm{NH}}\,\psi_{0}\left(t\right)
 +p_{\rm{H}}\int_{0}^{\infty}P_\kappa(\kappa)\,\psi_\kappa\left(t\right)d\kappa.
\end{eqnarray}

For the annealed case, since the hunters have constant density, perform independent random walks, and can simultaneously occupy the same site, 
$p_{\rm{NH}}$ results to be constant in time and can be expressed as $p_{\rm{NH}}= \left(\frac{m-1}{m}\right)^N$. This expression for the probability does not hold for low densities in the quenched case. The differences among these cases are discussed in Sec.~\ref{sec:numres}. 

We first consider a distribution $P_{\kappa}\left(\kappa\right)$ given by a Gamma distribution: 
\begin{equation}
\label{eq:Pkappa_gamma}
P_{\kappa}\left(\kappa\right)=  A \kappa^{-\sigma} e^{ \left [-\frac{b}{\kappa}\right]},
\end{equation}
with $A=\frac{b^{\sigma-1}}{\Gamma\left [\sigma-1\right ]} $, which has the same functional dependence of the distribution ($\chi^2$) one obtains by considering the inverse rate (i.e., the diffusivity) as the sum of several squared Gaussian random variables arising out of many microscopic degrees of freedom~\cite{2001BeckPRE}.

By performing a transformation of variable $D=1/\kappa$, we can solve the integral in Eq.~(\ref{Eq:Waitingtime})
\begin{equation}
 \frac{A}{\tau}  \int_0^\infty D^{\sigma-1} e^{\left [-Db\right ]} e^{\left [-\frac{Dt}{\tau} \right ]}   dD=\frac{b^{\sigma-1}}{\sigma}\left (b+\frac{t}{\tau}\right )^{-\sigma}.
\end{equation}
Thus the waiting time distribution shows a heavy-tail for $1<\sigma<2$. 

Another possible scenario is represented by the situation in which interactions with a hunter can only slow down the diffusion of the prey, thus corresponding to a distribution with a minimum value for $\kappa$, such as
\begin{equation}
\label{eq:Pkappa}
P_{\kappa}\left(\kappa\right)=(\sigma-1) \cdot \theta\left[ \kappa-1 \right] ~ \kappa^{-\sigma} 
\end{equation}
where $\theta\left[ \cdot \right]$  represents the Heaviside step function. 

In this case, the integral in Eq.~(\ref{Eq:Waitingtime}) gives 
\begin{equation}
\frac{1}{\tau}\int_{0}^{1} D^{\sigma-1}\,e^{\left[-\frac{D\, t}{\tau}  \right]}d\,D=\frac{1}{\tau}\left(\frac{t}{\tau}\right)^{-\sigma} \left(\Gamma[\sigma] - \Gamma[\sigma, t/\tau]\right),
\end{equation}
where $\Gamma[\sigma, t/\tau]=\int_{t/\tau}^\infty r^{\sigma-1}\exp[-r] d\,r $ is the upper incomplete Gamma function. 

This function converges exponentially to zero as $t\to\infty$. Therefore, in the long-time limit, it has the same behavior as the first term of Eq.~(\ref{Eq:Waitingtime}) and they can be both neglected.

It follows that the waiting time distribution for   $t\to\infty$ can be approximated as
\begin{eqnarray}
\label{Eq:Waitingtimedistribution}
	& \psi\left(t\right)\approx p_{\rm{H}}\frac{\Gamma[\sigma] }{\tau}\left(\frac{t}{\tau}\right)^{-\sigma}=\frac 1{\tilde \tau} \left(\frac t{\tilde \tau}\right)^{-\sigma} ,
\end{eqnarray}
where $(\frac{\tilde \tau}{\tau})^{\sigma-1}= p_H \Gamma[\sigma]$.

In order to prove that this distribution leads to a subdiffusive behavior, 
we calculate the eMSD $M_2(t)=\langle x^{2}\left(t\right)\rangle$, where $x$ is the distance of the prey from its initial position 
and $\langle\cdot\rangle$ represents ensemble average over multiple trajectories. 
Although the following calculations refer to the rate distribution in Eq.~(\ref{eq:Pkappa}), qualitatively similar results are obtained for any distribution as in Eq.~(\ref{eq:Pkappageneral}).

The Laplace transform of the eMSD, $M_2(s)$, can be obtained as
\begin{equation}
M_2(s)=\frac{ \psi(s)}{s\left[1- \psi(s) \right ]}\langle \ell^2 \rangle, \label{eq:m2s}
\end{equation}
where $\langle \ell^2 \rangle=1$ because the steps are taken on a regular lattice. 
According to a Tauberian theorem~\cite{book_klafter}, the Laplace transform of  Eq.~(\ref{Eq:Waitingtimedistribution}), once correctly normalized [i.e., $ \psi(s\rightarrow0)=1$], gives  
$\psi(s)\approx 1-\tilde \tau^\alpha s^\alpha$ for $\alpha=\sigma-1$ and $1<\sigma < 2$ . From Eq.~(\ref{eq:m2s}), it follows that
the eMSD at small $s$ can be expressed as $M_2(s)\approx \tilde \tau^{-\alpha} s^{-(\alpha+1)}$. As the behavior at small $s$ determines the one at large $t$, we find  
\begin{equation}
 M_2(t) = \mathrm{L}^{-1}\!\!\left[ M_2(s)\right]\!(t) \approx  \frac{1}{\Gamma(\sigma)}\!\left(\frac{t}{\tilde \tau} \right )^{\alpha}\!\!.
 \label{eq:eMSD}
\end{equation} 
We note that for   $\sigma\ge 2$  one gets $\alpha = 1$. Thus  $M_2(s) \propto 1/s$ and $M_2(t)$ grows linearly for $t\to \infty$.

It is worth recalling here that the coefficient $\tilde \tau$ depends through $p_{\mathrm{H}}$ on both the number of hunters $N$ and of sites $m$, 
\begin{equation}
\tilde \tau=\tilde \tau(N,m)\propto \left[1-\left(\frac{m-1}m\right)^N\right]^{\tfrac 1{\alpha}}.
\label{eq:tau}
\end{equation} 
For dilute systems (large $m$), one can approximate the probability of hitting a hunter as $p_{\mathrm{H}}\approx N/m$. This expression explicitly displays the dependence of the eMSD on the density of hunters $ \rho=N/m$, which can be verified both numerically and experimentally, thus providing a relevant experimental observable to test the model. We would like to point out that Eq.~(\ref{eq:tau}) is obtained for the case of annealed disorder.  For the quenched case, Eq.~(\ref{eq:tau}) still holds assuming that there is a negligible probability that the prey encounters the same hunter more than once before hitting a large number of other hunters. As we will discuss in Sec.~\ref{sec:numres}, this happens for large enough hunter densities. 
   
\begin{figure}
\includegraphics[width=0.95\columnwidth]{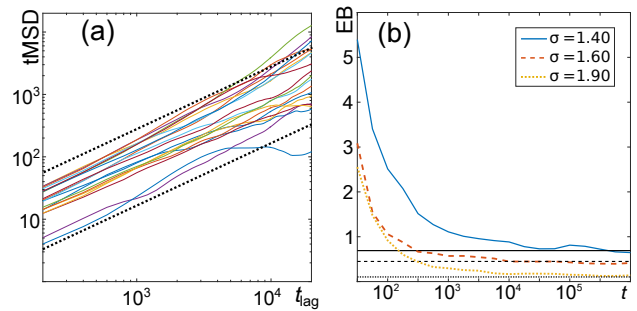}
\caption{(a) Time averaged mean squared displacement obtained for 20 representative prey trajectories with $\sigma=1.8$,  $N=6$ and $L=20$. 
The curves show a linear behavior but large scattering of their amplitude at all time lags, as expected for weak nonergodic behavior. 
Dashed lines correspond to linear behavior and are meant as a guide to the eye.  
(b) Ergodicity breaking (EB) parameter calculated as a function of total time of measurement $t$ for several values of $\sigma$ and $N=10$. 
At large $t$ all the curves asymptotically tend to the value predicted by Eq.~(\ref{eq:EB_value}) 
[horizontal black lines, line types correspond to different $\sigma$], 
as expected for a nonergodic behavior. 
As $\sigma$ is reduced, the system departs more from ergodicity and the EB shows larger asymptotic value.
\label{fig2}    }
\end{figure}

The presence of nonergodicity implies that time averages -- such as the time-averaged mean square displacement (tMSD),
\begin{equation*}
\overline{\delta^2}(t,t_{\mathrm{lag}})=\frac{\int_0^{t-t_{\mathrm{lag}}} \left[x(t'+t_{\mathrm{lag}})-x(t')\right]^2dt'}{t-t_{\mathrm{lag}}},
\end{equation*}
calculated for different trajectories -- remain random variables at all time lags.
To give a quantitative measure of the nonergodicity,  the ergodicity breaking (EB) parameter was introduced as~\cite{2008HePRL}
\begin{equation}
\label{eq:EB}
 \mathrm{EB}=\lim_{t\to\infty}\frac{\langle(\overline{\delta^2})^2\rangle-\langle\overline{\delta^2}\rangle^2}{\langle\overline{\delta^2}\rangle^2}.
\end{equation}
In the case of an ergodic process the EB parameter should be zero, while $\rm{EB}>0$ indicates nonergodicity. 
For a  heavy tailed CTRW, where the waiting times are distributed according to a power law,  the EB parameter  is given by~\cite{2008HePRL}
\begin{equation}
 \mathrm{EB}=\frac{2\Gamma^2[\sigma]}{\Gamma[2\sigma-1]}-1.
 \label{eq:EB_value}
\end{equation}
This result also applies to the case described here, as one can rewrite  Eq.~(\ref{Eq:Waitingtimedistribution}) as $\psi(t)\approx t^{-(1+\alpha)}/|\Gamma(-\alpha)|$.

Finally, we discuss the deviation from Gaussianity of the propagator $P(x,t)$, 
which describes the probability of finding the prey at position $x$ at time $t$~\cite{2006HoflingPRL,2013HoflingRPP,2015MerozPR,2015GhoshPCCP,2016ThorneyworkSM}. 
A standard measurement of such deviation is given by the functional~\cite{1964RahmanPR}
\begin{equation}
 \vartheta(P(x,t)) =\frac{\langle  x^4\rangle}{a(d)\langle  x^2\rangle^2}-1, 
 \end{equation}
 which is usually called non-Gaussianity parameter of $P(x,t)$ and it is well defined 
 for any number of spatial dimensions $d$. The $d$-dependent constant $a(d)=1+2/d$ corresponds to
 the ratio $\tfrac {\langle  x^4\rangle}{\langle  x^2\rangle^2}$ computed with a Gaussian propagator in $d$ dimensions and derived in the Appendix~\ref{sec:app}.
 Thus, $\vartheta$ is defined to be zero for a Gaussian process and $\vartheta(P(x,t))=0$ is a necessary but not sufficient condition for $P(x,t)$ to be Gaussian.  
 It is worth noticing that, for a generic process, $\vartheta$ is time dependent.  
 Here, we are interested in calculating the value of $\vartheta$ at late time for a 1D CTRW, for which $a(d=1)=3$. 
We assume a CTRW with the waiting time distribution given in Eq.~( \ref{Eq:Waitingtimedistribution}) 
 and with a generic distribution $p(x)$ of step size. In our particular case the steps have all the same length one, so $p(x)=\delta(x-1)$. 
 In order to obtain $\vartheta$, we need to calculate 
 $M_4(t) = \langle x^4(t)\rangle $. As we are interested in the late time behavior of $M_4(t)$, it is convenient to compute
 its Fourier-Laplace transform $M_4(s)$ at small $s$.    
  From the Fourier-Laplace transformed propagator, given by
 \begin{equation}
  P(k,s)=\frac{1-\psi(s)}{s}\frac{1}{1-\lambda(k)\psi(s)},  
 \end{equation}
 with $\lambda(k)=\int \exp[-i k x] p(x)ds $, all moments can be obtained as
 \begin{equation}
  M_n(s)=(-i)^n \left.\frac{d^n P(k,s)}{dk^n} \right|_{k=0}.
 \end{equation}
 It thus follows that
 \begin{equation}
  M_4(s)=\frac{\psi(s)}{s}\left(\frac{\langle\ell^4\rangle}{1-\psi(s)}+\frac{6\langle\ell^2\rangle\psi(s)}{(1-\psi(s))^2}\right). 
 \end{equation}
Similarly as discussed above, by considering  $\psi(s)\approx (1-\tilde \tau^\alpha s^\alpha)$, we get at leading order
$M_4(s)\approx 6 \langle\ell^2\rangle^2 s^{-1 - 2 \alpha} \tau^{-2 \alpha},$
and by comparison with \eqref{eq:eMSD}, we obtain      
 \begin{equation}
 M_4(t) \approx 6 M_2(t)^2, \ \ \ \ \ t\gg\tilde \tau,
\end{equation} 
for which $\lim_{t\to \infty} \vartheta (t) = 1$. Thus, our process displays a non-Gaussian propagator, as expected for being a generical CTRW process. 
%
%
\section{Numerical results} 
\label{sec:numres}
We simulated the master equation associated to the system of a prey and $N$ independent hunters by means of a Monte Carlo method in 1D. We take $\tau=1$ and thus we consider {\it long times} as $t$ larger than 1 in three or four orders of magnitude. Unless otherwise stated, the calculations are performed for the annealed binding rates case.  The numerically-calculated waiting time distribution is represented in Fig.~\ref{fig1} (a) together with the analytical prediction $ \psi(t)$ in Eq.~(\ref{Eq:Waitingtime}), showing an excellent agreement. In panel (b) of this figure, we also plot the curves obtained by the calculation of eMSD, showing the asymptotic subdiffusive behavior  $\sim t^\alpha$ for $1<\sigma<2$, as predicted by the theory. 

Consistently with the model prediction, the representative tMSD curves shown in Fig.~\ref{fig2}(a) display a linear behavior but large scattering of their values at all time lags, confirming that time averages are random and thus lead to nonergodicity. To quantitatively corroborate this observation, we calculated EB with Eq.~(\ref{eq:EB}). In Fig.~\ref{fig2}(b),  we show that  the value of EB converges to the value theoretically predicted by Eq.~(\ref{eq:EB_value}) for $1<\sigma<2$. As $\sigma$ is decreased from 2 towards 1 the system becomes more nonergodic (larger EB).

\begin{figure}
\includegraphics[width=0.95\columnwidth]{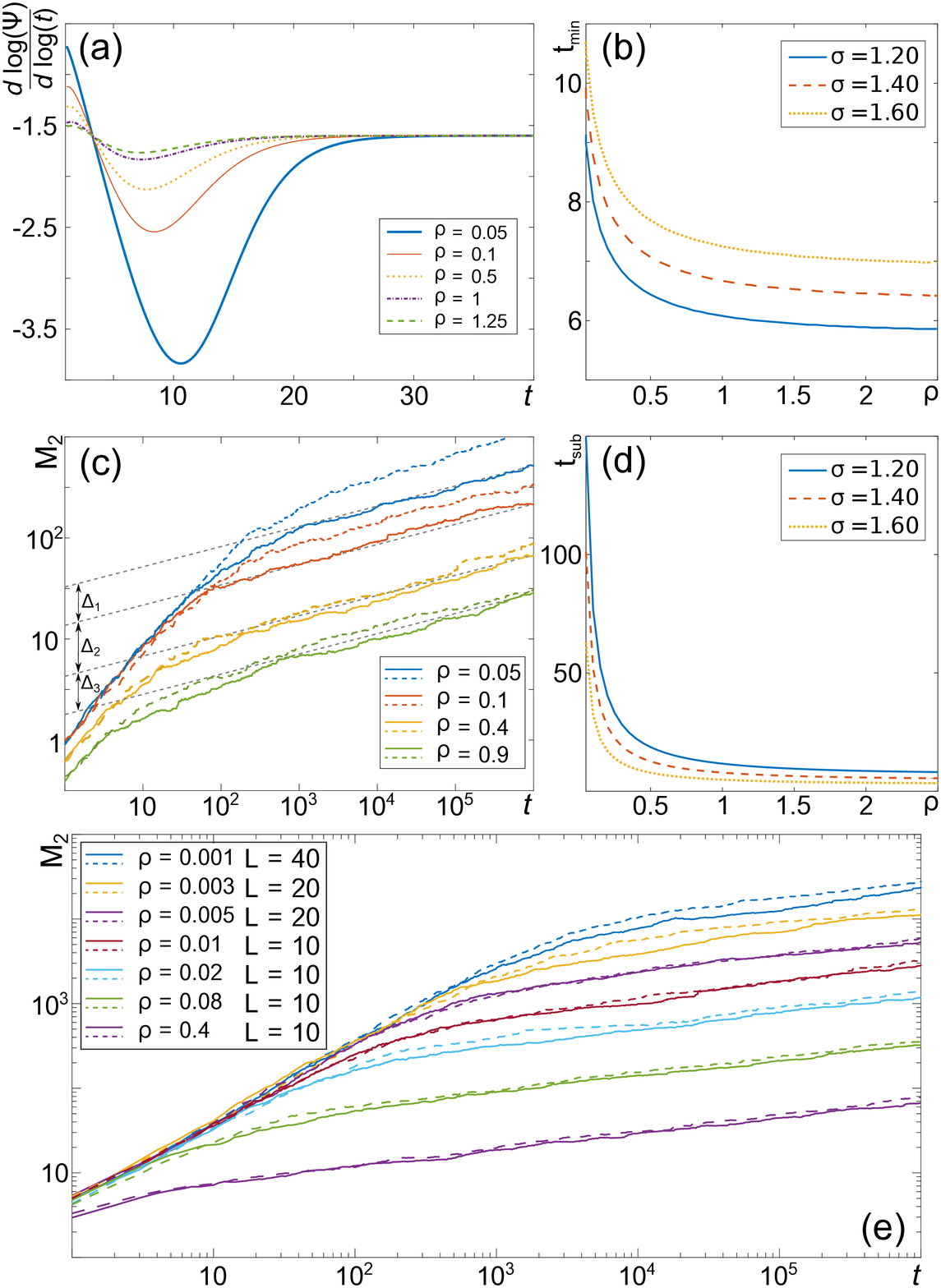}
\caption{(a) Logarithmic derivative of the distribution of waiting times as in Eq.~(\ref{Eq:Waitingtime}) for $\sigma=1.6$ and different values of the density of hunters $\rho$ ($L=20$). For large $t$, the derivative tends to the expected value of $-\sigma$ [corresponding to the exponent of Fig.~\ref{fig1}(a) at long times]. (b) Time $t_{\rm{min}}$ at which the minima of the curve in panel (a) are reached, as a function of the density. The larger the density, the quicker the logarithmic derivative of the waiting time distribution tends to its asymptotic value $\sigma$ (shorter $t_{\rm{min}}$) . (c) eMSD for $\sigma=1.2$ in 1D as obtained at different densities for annealed (continuous lines) and quenched disorder (dashed lines). The long-time scaling exponent of the eMSD is independent of the density and the type of disorder. In the asymptotic regime, curves obtained for different densities and type of disorder are separated by a distance $\Delta$. (d) Analytically calculated time $t_{\rm{sub}}$ at which the subdiffusive behavior occurs as a function of $\rho$.  (e) eMSD for $\sigma=1.2$ in two dimensions for different densities and annealed (continuous lines) or quenched disorder (dashed lines). 
\label{fig3}    }
\end{figure}

Next, we obtain that the long time properties of the system, like the exponent $\alpha$ quantifying the subdiffusive behavior and the value of the EB parameter,  do not depend on the density of hunters $\rho$. However, as  $\rho$ is increased, the weight $p_{\rm{H}}$ also increases, thus the second term of the probability distribution Eq.~(\ref{Eq:Waitingtime}) becomes dominant with respect to the first one at shorter times. To exemplify this behavior, in Fig.~\ref{fig3}(a) we plot the logarithmic derivative of the distribution of waiting times Eq.~(\ref{Eq:Waitingtime}), for different densities. All the curves cross at a value equal to $\sigma$, which can be analytically calculated to occur at $t=\sigma\tau$. In addition, all the curves show a qualitatively similar behavior, decreasing to reach a minimum and then increasing towards the asymptotic value $\sigma$. In Fig.~\ref{fig3}(b), we plot the time $t_{min}$ at which this minimum is reached as a function of the density, showing that -- as $\rho$ increases -- the time at which the curve approaches its asymptotic value decreases. In Fig.~\ref{fig3}(c) we plot the eMSD for different densities and $\sigma=1.2$. In the same panel, we also show eMSD obtained for annealed and quenched disorder of binding rates. In all the cases, the curves asymptotically tend to a power law with exponent $\alpha\simeq0.2$, as expected. However, for low densities differences arise between the two models, with the curves corresponding to the quenched disorder reaching the asymptotic regime at larger times and thus showing a higher offset with respect to the annealed case. This difference is caused by the fact that, for quenched rates and low densities, the probability of meeting twice the same hunter before hitting a large number of times other hunters is not negligible. This implies that it takes more time for the prey to sample the disorder due to the hunters. In 1D and high density, this probability becomes negligibly small; thus the difference between the quenched and annealed case is very small.  In two-dimensional lattices, the difference between the quenched and annealed case appears at even smaller densities with respect to 1D, as shown in Fig.~\ref{fig3}(e).
Once the asymptotic behavior is reached, the distance between eMSD curves obtained for different number of particles $N_i$ and $N_j$ and the same $m$ is given by $ \Delta=\log(p_{\mathrm{H}}(N_j,m)/p_{\mathrm{H}}(N_i,m))$, which reduces to $ \Delta=\log(N_j/N_i)$ in the dilute limit [see schematic representation in Fig.~\ref{fig3}(c)] . Again, for the annealed case and the quenched case at large density, we found a good agreement between this analytical prediction and the numerical calculation in 1D and 2D [see Figs.~\ref{fig3}(c) and (e)] For the quenched case, as the density is reduced, the calculations show deviation from this prediction.  

Finally,  due to the dependence of eMSD on $p_{\mathrm{H}}$, one can give a lower bound for the time required for the subdiffusive behavior to appear, $t_{\rm{sub}}$.  This lower bound can be obtained as the time at which the logarithm of the long-time power-law behavior [Eq.~(\ref{eq:eMSD})] and of the short-time linear behavior of the eMSD intersect, giving $t_{\rm{sub}}=\frac{10^{1-\alpha}}{p_{\rm{H}}\Gamma(\sigma)^2}$. 
From the plot of $t_{\rm{sub}}$ in Fig.~\ref{fig3}(d) -- evaluated for the annealed case --  one can notice that while its qualitative behavior is similar to $t_{\rm{min}}$ [Fig.~\ref{fig3}(b)],   $t_{\rm{sub}}$ is always larger than $t_{\rm{min}}$  as expected. Remarkably, for the same number of hunters and sites,  $t_{\rm{sub}}$ and $t_{\rm{min}}$ have inverse dependence to $\Gamma(\sigma)$. Increasing $\sigma$, the heavy tailed part of the waiting time distribution Eq.~(\ref{Eq:Waitingtime}) becomes dominant at larger times, while subdiffusion emerges at shorter times.

\section{Conclusions} The model discussed in this paper proposes a mechanistic explanation for the nonergodic subdiffusion observed in several biological systems~\cite{2006GoldingPRL, 2011JeonPRL, 2011WeigelPNAS, 2013TabeiPNAS, 2015ManzoPRX}, based on the occurrence of specific, transient interactions with a heterogeneous population of interacting partners.  The model displays many similarities with the long-time behavior of the heavy-tail CTRW~\cite{1975ScherPRB} and the patch model~\cite{2014MassignanPRL}. However, there are crucial differences that, in our opinion, make our model extremely suitable for the interpretation of experimental data of molecular diffusion on the cell membrane. First,  in contrast to the heavy-tail CTRW model, the waiting time distribution -- and thus the distribution of interaction time between the prey and a hunter -- is exponential as expected for a Markov process. Therefore, our model allows one to reproduce nonergodic behavior without requiring one to postulate time-dependent effects to justify power law distributions for the interactions or trapping time~\cite{2013Weigel}. 
For the CTRW with heavy-tail waiting time distribution, the presence of ``annealed" or ``quenched'' disorder produces different diffusion laws in $d=1,2$. This is a consequence of the fact that -- when the rate at a given lattice site is constant (``quenched'') -- correlations arise between successive arrival times~\cite{1990BouchaudPhysRep}. Our model can also be formulated in two versions, one in which a hunter picks a random rate every time it encounters a prey (annealed disorder), and one in which each hunter has a constant rate (quenched disorder). For both realizations of our model, the sampling of transition rates -- and therefore the disorder --  is highly dynamical as it occurs through the hunter-prey interaction mechanism, and thus induces low correlations which are smaller for large densities or large dimensionality. We find differences between the two realizations in 1D and low hunter densities and 2D for much smaller densities. However, in both cases we obtain the same scaling asymptotic behavior, to which our model exactly converges to in the high density limit ($\rho \to \infty$).

Our model also shows similarities with the long-time behavior of the time-annealed patch model~\cite{2014MassignanPRL} for the value of the exponent $\gamma=1$. However, in the patch model, the disorder is ubiquitous in space and/or time and the random walker continuously undergoes changes of diffusivity. In contrast, in our case the prey typically diffuses in a Brownian fashion with constant $D$, except for the local changes of diffusivity experienced for a limited amount of time upon interactions with the hunter. Therefore, we show that anomalous diffusion and nonergodicity can emerge even in the presence of partial disorder (small $\rho$).

The proposed framework relies on the assumption of a broad distribution of the diffusion rates (or diffusion coefficients) of the interacting partners. We consider this assumption rather reasonable since the hunters in our model might represent different chemical species and on the basis of broad diffusivity distributions reported for chemically identical cellular components~\cite{1997SaxtonBJ}. Moreover, our general requirements for the distribution of rates include the particular case in which the diffusivity is the sum of several squared Gaussian random variables, e.g. due to the presence of a large number of degrees of freedom~\cite{2001BeckPRE}.

An important feature of the model is the possibility of being experimentally tested, thus allowing one to distinguish its occurrence from other theoretical frameworks. This is nowadays technologically possible by means of multicolor single particle tracking techniques. As an example, in a dual color single particle tracking experiment it is possible to simultaneously follow the motion of two closely spaced particles with time resolution of few milliseconds and resolve their relative distance with a precision of the order of ~10 nm~\cite{2015ManzoROPP,2015TorrenoJPD}. Analogously to single particle tracking, these experiments provide trajectories from which the time- and ensemble-averaged MSD can be calculated, thus allowing one to test the appearance of nonergodicity. In addition, the technical advantages afforded by dual color tracking make it possible to experimentally verify the occurrence of interactions between diffusing species, measure the duration of such events, and check whether they affect the diffusivity of the particles involved~\cite{2011DLidke, 2012Kusumi, 2015TorrenoJPD}. These experiments can be carried out by labeling chemically identical components as well as different species, thus testing the formation of both homo- and hetero-oligomers. This technique has already been successfully used to study interactions of several membrane components~\cite{2011DLidke, 2012Kusumi, 2015TorrenoJPD}. 
In addition, other promising approaches to investigate interaction-dependent diffusion include hyper-spectral microscopy~\cite{2013CutlerPlosONE}, as well as the combination of single particle tracking with recent methods based on advanced statistical tools~\cite{2014MassonBJ, 2016Masson} and on the spatiotemporal analysis of fluorescence fluctuations~\cite{2016Gratton}, which have been shown to provide a wealth of information into dynamic molecular processes of biological relevance. 

We would like to further stress that, while these experimental strategies allow one to discriminate on the occurrence of our theory in a specific system, the model allows one to directly calculate microscopic parameters of the system under investigation.  Indeed, the timescale for the onset of subdiffusion in the eMSD curve provides an estimation of the average density of hunters, thus quantifying the level of crowding experienced by the prey. In addition, the scaling exponent of the eMSD is a proxy for the degree of heterogeneity of the environment. However, for a robust determination of these parameters one needs to collect  a sufficient number of eMSD data points over the appropriate timescale. For example, in order to precisely extract the time at which subdiffusion arises, one needs to collect  a sufficient number of eMSD data points spanning over at least two orders of magnitude centered around such a timescale. In typical SPT experiments, this range is bounded by the time-resolution and the trajectory duration~\cite{2015ManzoROPP}. The time resolution (i.e. the inverse of the recording frame rate), besides setting the shortest eMSD time point, also determines the lag between successive points and thus the number of data points within the measured range. The maximum trajectory length is instead ultimately limited by the photon budget of the fluorescent emitter. Therefore, although it is desirable to collect a large number of photons in each frame in order for the precise localization of the particle~\cite{2015ManzoROPP}, this would limit either the number of points or the maximum duration of the trajectory. Therefore, the experimental conditions must be finely tuned in order to obtain the best tradeoff between tracking precision, time resolution and trajectory length. Although it is currently possible to obtain eMSD with hundreds of data points between a few milliseconds to tens of seconds, new strategies have the potential to push these bounds even further~\cite{2016Balzarotti}. In this scenario, we think that our model might be a useful tool to investigate anomalous transport and its implications, while providing an alternative interpretation to the causes of non-ergodic subdiffusion.

\section{Acknowledgements} We acknowledge discussions with Gerald John Lapeyre, Jr. and useful comments to the first arXiv version from Eli Barkai. The authors acknowledge financial support from the Spanish Ministry of Economy and Competitiveness, through the ``Severo Ochoa'' Programme for Centres of Excellence in R\&D (SEV-2015-0522); Fundaci\'o Cellex; Generalitat de Catalunya (Grants No. 2009 SGR 597 and No. 2014 SGR 874), the European Commission [FP7-ICT-2011-7, Grant No. 288263; SIQS (FP7-ICT-2011-9, Grant No. 600645);  EU STREP QUIC (H2020-FETPROACT-2014 No. 641122); EQuaM (FP7/2007--2013 Grant No 323714)]; the HFSP (Grant RGP0027/2012); OSYRIS (ERC-2013-AdG Grant No. 339106); the Spanish Ministry of Science and Innovation (Grants FOQUS FIS2013-46768-P, FISICATEAMO FIS2016-79508-P,  and No. FIS2014-56107-R), and the CERCA Programme/Generalitat de Catalunya. C.M. acknowledges funding from the Spanish Ministry of Economy and Competitiveness and the European Social Fund (ESF) through the Ram\'on y Cajal program 2015 (No. RYC-2015-17896).

\appendix

\section{Gaussianity of the propagator}
\label{sec:app}

Let us assume a Gaussian propagator with the normalized form 
\begin{equation}
P({\bf x},t)= (4\pi\rho(t))^{-d/2}\exp[-x^2/4\rho(t)],
\end{equation}
where ${\bf x}$ is the displacement in a $d$-dimensional space $\mathbb{R}^d$, $x$ its modulus, and $\rho(t)$ is the variance and has dimensions of a length to the square. 
All momenta are calculated as
\begin{align}
 \langle x^{2n}\rangle& =\int_{\mathbb{R}^d} d^d {\bf x} \,x^{2n}P({\bf x},t)\nonumber\\
 &=\frac{(4\rho)^n}{(4\pi\rho)^{d/2}}\frac{\partial^n}{\partial k^n}\int_{\mathbb{R}^d}\exp\left[-k\,x^2/4\rho\right]\Bigg|_{k=1}\nonumber\\
 &=\frac{(4\rho)^n}{(4\pi\rho)^{d/2}}\frac{\partial^n}{\partial k^n}\left[\frac{4\pi\rho}{k}\right]^{d/2}\Bigg|_{k=1}\nonumber\\
 &=(-4\rho)^n \, \frac{d}{2}\, \frac{d+2}{2}\cdots \frac{d+2(n-1)}{2}.\label{eq:allmomenta}
\end{align}
Therefore, for a Gaussian propagator, the expression~(\ref{eq:allmomenta}) can be used to show that the ratio between the fourth moment and the square of the second moment is given by
\begin{equation}
 a(d) \equiv \frac{ \langle x^4\rangle}{\langle x^2\rangle^2}=\frac{d(d+2)}{d^2} = 1 + \frac 2d. 
\end{equation}

\end{document}